\documentstyle[12pt]{article}

\newcommand{\be}{\begin{equation}} 
\newcommand{\ee}{\end{equation}} 
\newcommand{\bea}{\begin{eqnarray}} 
\newcommand{\eea}{\end{eqnarray}} 
\newcommand{\nn}{\nonumber \\} 
\newcommand{\p}[1]{(\ref{#1})} 
  
\newcommand\al{\alpha}
\newcommand\bt{\beta}
\newcommand\de{\delta}
\newcommand\ga{\gamma}
\newcommand\ep{\epsilon}

\begin{document}
\begin{titlepage}
\begin{flushright}
hep-th/9811244 \\
November 28, 1998
\end{flushright}

\vskip2cm
\begin{center}
{\large\bf Partial breaking of $N=1,\;  D=10$ supersymmetry} 
\end{center}

\vskip1cm
\centerline{\bf S. Bellucci,}

\vskip.1cm
\centerline{\it INFN - Laboratori Nazionali di Frascati,}
\centerline{\it P.O. Box 13, I-00044 Frascati, Italy}
\centerline{\it e-mail: bellucci@lnf.infn.it}

\vskip.3cm
\centerline{\bf E. Ivanov, S. Krivonos}

\vskip.1cm
\centerline{\it Bogoliubov Laboratory of Theoretical Physics, JINR,}
\centerline{\it 141 980, Dubna, Moscow Region, Russian Federation}
\centerline{\it e-mail: eivanov@thsun1.jinr.ru, krivonos@thsun1.jinr.ru}

\vskip.8cm
\begin{abstract}
\noindent We describe the spontaneous partial breaking 
of $N=1,\;D=10$ supersymmetry to $N=(1,0),\;d=6$ and 
its dimensionally-reduced 
versions in the framework of nonlinear realizations. 
The basic Goldstone superfield is  
$N=(1,0),\;d=6$ hypermultiplet superfield satisfying a nonlinear 
generalization of the standard hypermultiplet constraint. 
We interpret the generalized constraint as the manifestly worldvolume
supersymmetric form of equations of motion of the type I 
super $5$-brane in $D=10$. The related issues 
we address are a possible existence 
of brane extension of off-shell hypermultiplet actions, 
the possibility to utilize vector $N=(1,0),\; d=6$ supermultiplet 
as the Goldstone one, and the description of $1/4$ breaking of 
$N=1,\;D=11$ supersymmetry. 
\end{abstract}

\end{titlepage}

\noindent{\bf 1. Introduction.} 
The description of partial breaking of global 
supersymmetries (PBGS) within the coset approach \cite{1} - \cite{3}
received much attention \cite{bw} - \cite{gpr}. 
Its characteristic feature is that the Goldstone fermionic fields
associated with the broken supertranslation generators \cite{vak} 
come out as components of Goldstone multiplets of unbroken SUSY. 

The study of different patterns of PBGS in refs. 
\cite{bw} - \cite{gpr} revealed a few peculiarities 
of such theories. As applied to the most elaborated case of 
the $1/2$ partial breaking of $N=2,\; D=4$ SUSY, these are as follows. 
\begin{itemize}
\item There are several inequivalent $N=1$ Goldstone supermultiplets 
related with the partial breaking $N=2 \rightarrow N=1$: chiral \cite{bg1}, 
vector \cite{bg2} and tensor ones \cite{{bg3},{gpr}}. 
These options correspond to different theories. 

\item The $N=1$ superfield Goldstone actions can be 
treated as gauge-fixed, manifestly worldvolume 
supersymmetric forms of the actions of some 
BPS superbranes, along the line of refs. \cite{hp,hlp}. 
The $N=1$ chiral Goldstone superfield action 
is recognized as that of the Type I super 3-brane in a flat $D=6$ 
background \cite{hlp}. The $N=1$ vector 
Goldstone multiplet action describes a super D3-brane and yields 
the Born-Infeld (BI) action for the gauge field. In all cases 
the no-go theorem of \cite{witten} is evaded by the general argument 
of \cite{hp}.  

\item In accord with the general features of nonlinear realizations, 
one can make different $N=1$ matter actions $N=2$ supersymmetric 
by coupling them to Goldstone superfields. 
\end{itemize} 

The actions presented in \cite{bg1} - \cite{bg3} are nonlinear, 
``brane'' generalizations of familiar off-shell $N=1$ superfield 
actions. On the other hand, theories with {\it linearly} 
realized $N=2,\;  d=4$ SUSY admit a good off-shell description, e.g.  
in harmonic $N=2$ superspace \cite{gikos}. It is natural to ask 
whether some of them can be promoted 
to those with a nonlinearly realized higher SUSY, say $N=4$ SUSY, 
by constructing 
the formalism of partial breaking of this higher SUSY down to 
$N=2$ and identifying some $N=2$ superfields as 
the Goldstone ones accompanying this breakdown. Related questions are as to
what kind of superbranes could be associated with such theories, whether 
a brane generalization of the harmonic analyticity \cite{gikos} 
underlying ordinary $N=2$ theories exists, how many different Goldstone 
$N=2$ superfields are possible, etc.  

In this letter we partly answer these questions. 
We show that  
the partial breaking of 
$N=1,\; D=10$ SUSY (amounting to properly central-charge extended 
$N=4$ SUSY in $d=4$ or $N=(1,1)$ SUSY in $d=6$) down to $N=(1,0),\; d=6$ SUSY 
picks out $d=6$ hypermultiplet as the 
basic Goldstone superfield.  
Using the coset space techniques, 
we find a covariant nonlinear generalization of 
the standard hypermultiplet constraint in $N=(1,0),\; d=6$ 
superspace \cite{fs}. 
We argue that the generalized 
constraint encodes a gauge-fixed form of the equations of motion of the 
super 5-brane in $D=10$ with manifest worldvolume 
$N=(1,0),\; d=6$ SUSY. 
We give an evidence for the existence of brane extensions 
of the harmonic analyticity and  
off-shell hypermultiplet actions. Our relations admit the dimensional 
reduction by the worldvolume bosonic dimension up to 
the extreme $N=8, \; d=1$ case corresponding to a superparticle in 
$D=5$. We elaborate on this simple case in more detail. Finally, we briefly 
discuss some related questions, in particular, a possibility to  
apply the PBGS approach to $N=1,\;D=11$ SUSY.     

\vspace{0.3cm}
\noindent{\bf 2. $N=1,\; D=10$ Poincar\'e superalgebra in the $d=6$ notation.}  
{}From the $d=6$ viewpoint 
the $N=1,\; D=10$ SUSY algebra is a central-charge 
extended $N=(1,1)$
Poincar\'e superalgebra:   
\be 
N=1,\; D=10\;\;\;\; SUSY \qquad \propto \quad
\left\{ Q^i_\alpha, P_{\alpha\beta}, S^{\beta a}, Z^{ia}
\right\}~, \label{setsus}
\ee
where 
$$
\alpha, \beta = 1,...,4~, \quad i = 1,2~, \quad a = 1,2 
$$ 
are, respectively, the $d=6$ spinor ($Spin(1,5)$) indices and the 
doublet indices of two 
commuting automorphism $SU(2)$ groups realized on the spinor $Q$ and $S$
generators (see \cite{hst} - \cite{bz} for the $d=6$ spinor notation). 
The basic anticommutation relations read 
\be
\left\{ Q_{\al}^i,Q_{\bt}^j\right\}=\ep^{ij}P_{\al\bt}\; ,\quad
\left\{ Q_{\al}^i,S^{a\bt}\right\} = \de_{\al}^{\bt}Z^{ia}\; , \quad
\left\{ S^{a\al},S^{b\bt}\right\} = \ep^{ab}P^{\al\bt} \;. \label{susy}
\ee
The $d=6$ translation generator
$P_{\alpha\beta} = -P_{\beta\alpha} = 
{1\over 2}\epsilon_{\alpha\beta\rho\lambda}P^{\rho\lambda}$, together 
with the "semi-central charge" generator $Z^{ia}$, form 
the $D=10$ translation generator. 

To the set \p{setsus} one should add the generators of 
the $D=10$ Lorentz group $SO(1,9)$ 
\be 
SO(1,9) \qquad \propto \quad \left\{ M_{\al\bt\;\ga\de},\;\;
T^{ij}, \;\; T^{ab}, \;\; K_{ia}^{\al\bt}\right\}~. \label{so19ge}
\ee
The generators $M$ and $T$ generate mutually commuting $d=6$ Lorentz
group $SO(1,5)$ and the automorphism (or $R$-symmetry) 
group $SO(4)\sim SU(2)\times SU(2)$, the generators 
$K$ belong to the coset $SO(1,9)/SO(1,5)\times SO(4)$.  
 
\vspace{0.3cm}
\noindent{\bf 3. Coset space routine.} We are going to 
construct a nonlinear realization of $N=1,\; D=10$ SUSY (together with 
the $D=10$ Lorentz group), such that $N=(1,0),\; d=6$ SUSY remains unbroken. 
Thus we choose the vacuum stability subgroup to be 
\be
H \quad \propto \quad \left\{Q^i_\alpha, P_{\alpha\beta}, T^{ij}, T^{ab}, 
M_{\al\bt\;\ga\de}
\right\}~. \label{Hsub} 
\ee
We put the generators $Q^i_\alpha, 
P_{\alpha\beta}$ into the coset and associate with them as the coset
parameters the coordinates of $N=(1,0),\; d=6$ superspace 
\be 
Q^i_\alpha \Rightarrow \theta^\alpha_i~, \quad P_{\alpha\beta} 
\Rightarrow x^{\alpha\beta}~. \label{SS}
\ee
The remaining coset generators, $S^{\alpha a}, Z^{ia},
K^{ia}_{\alpha\beta}$, correspond to genuine spontaneously broken 
symmetries. The corresponding coset parameters are Goldstone superfields  
\be
S^{\alpha a} \Rightarrow \Psi_{\alpha a}(x, \theta)~, \quad 
Z^{ia} \Rightarrow q_{ia}(x,\theta)~, \quad K^{ia}_{\alpha\beta} 
\Rightarrow \Lambda^{\alpha\beta}_{ia}(x,\theta)~. \label{Golddef}
\ee
An element $g$ of the coset 
space $G/\tilde{H}$, where $G$ is the full 
supergroup of $N=1,\; D=10$ SUSY (including $SO(1,9)$) 
and $\tilde{H} = SO(1,5)\times SO(4)$, reads
\be
g=e^{x^{\al\bt}P_{\al\bt}}e^{\theta^{\al}_iQ_{\al}^i}e^{q_{ia}Z^{ia}}
  e^{\Psi_{a\al}S^{a\al}}e^{\Lambda_{ia}^{\al\bt}K_{\al\bt}^{ia}} \;.
\label{g}
\ee
Acting on \p{g} from the left by different elements of $G$ with 
constant parameters, one determines the transformation properties 
of the coset parameters.   

Unbroken supersymmetry $(g_0=\mbox{exp }(a^{\al\bt}P_{\al\bt}+
  \eta_i^{\al}Q_{\al}^i ))$:
\be\label{susy1}
\de x^{\al\bt}=a^{\al\bt}+\frac{1}{4}\left(
\eta^{i\al}\theta_i^{\bt} - \eta^{i\bt}\theta_i^{\al} \right),\quad
\de \theta_{i}^{\al}=\eta_i^{\al}\; .
\ee

Broken supersymmetry $(g_0=\mbox{exp }(\eta_{a\al}S^{a\al}))$:
\be\label{susy2}
\de x^{\al\bt}=\frac{1}{4}\ep^{\al\bt\ga\de}\eta_{\ga}^a\Psi_{a\de},\quad
\de q_{ia}=-\eta_{a\al}\theta_i^{\al},\quad
\de\Psi_{a\al}=\eta_{a\al}\; .
\ee

Broken $Z$-translations $(g_0 = \mbox{exp}(c_{ia}Z^{ia}))$: 
\be \label{Ztr}
\de q^{ia} = c^{ia}~.
\ee

The form of broken $K$ transformations is irrelevant 
for our consideration. The subgroup $\tilde{H}$ 
is realized as rotations of the $SO(1,5)$ spinor and $SU(2)$ doublet indices.

We see that $N=1,\; D=10$ supergroup as a whole admits a realization on 
the coordinates of $N=(1,0),\; d=6$ superspace and Goldstone superfields
living on this superspace. 

The next step is the construction of the left-covariant Cartan 1-forms:
\be
g^{-1}d g = \Omega_Q + \Omega_P + \Omega_Z + \Omega_S + \Omega_K +
\Omega_{\tilde H}~, 
\ee 
where the subscripts denote the relevant generators. 
We shall actually need only the form $\Omega_Z$    
\be
\Omega_Z \equiv \Omega_Z^{ia}\;Z_{ia} = \left[ \left( \mbox{ch } 
\sqrt{\varphi}\right)^{ia}_{jb}
   d{\hat q}^{jb}+\left( \frac{\mbox{sh }\sqrt{\varphi}}{\sqrt{\varphi}}
\right)^{ia}_{jb}2\Lambda^{jb\mu\nu}d{\hat x}_{\mu\nu} \right]Z_{ia}~,
\label{cforms}
\ee
\bea
d{\hat x}^{\al\bt}& = & dx^{\al\bt}-\frac{1}{4}\theta^{i\al}d\theta_i^{\bt}
  +\frac{1}{4}\theta^{i\bt}d\theta_i^{\al}-
  \frac{1}{4}\epsilon^{\al\bt\mu\nu}\Psi^a_{\mu}d\Psi_{a\nu}~, \nn
d{\hat q}_{ia}&=& dq_{ia}+\Psi_{a\al}d\theta_i^{\al}~, \quad 
\varphi^{ia}_{jb} \equiv  2 \Lambda^{ia\mu\nu}\Lambda_{jb\mu\nu}~. \label{def2}
\eea

\vspace{0.3cm}
\noindent{\bf 4. Inverse Higgs constraints and dynamical equation.}
By construction, the Cartan form \p{cforms} is covariant under all
transformations of $G$ realized as left shifts of $g$. The Goldstone 
superfields $\Lambda^{\alpha\beta}_{kb} $ and $\Psi_{\alpha a}$ appear 
inside it {\it linearly} and so can be covariantly eliminated by the 
inverse Higgs procedure \cite{invh}. This is
achieved by imposing the manifestly covariant constraint
\be
\Omega_Z = 0~. \label{basconstr}
\ee
It amounts to the following set of equations 
\be 
\widetilde{\Lambda}^{ia}_{\rho\sigma} 
\equiv
    -2\left( \frac{\mbox{th }\sqrt{\varphi}}{\sqrt{\varphi}}
\right)^{ia}_{jb} \Lambda^{jb}_{\rho\sigma}
= (E^{-1})_{\rho\sigma}^{\mu\nu}
\;\partial_{\mu\nu}q^{ia} \equiv \nabla_{\rho\sigma} q^{ia}~, \qquad
\Psi_{a\beta} = {1\over 2}\;\nabla_\beta^k \;q_{ka}~, \label{psieq} 
\ee
\be
\nabla^{(i}_\bt \;q^{k)a} = 0~. \label{basconstr2}
\ee
Here 
\bea
&&\nabla^k_\beta \equiv {\cal D}^k_\bt - 
{1\over 4}\;
\epsilon^{\rho\lambda\al\ga}(\Psi_{\al}^b {\cal D}_{\bt}^k \Psi_{b\ga})
\;\nabla_{\rho\lambda}~, 
\label{defnabl} \\ 
&&E^{\rho\sigma}_{\mu\nu} \equiv \frac{1}{2}
 \left(\de_{\mu}^{\rho}\de_{\nu}^{\sigma}  -
\de_{\mu}^{\sigma}\de_{\nu}^{\rho}
-\frac{1}{2}
\epsilon^{\rho\sigma\al\bt}\Psi_{\al}^b \partial_{\mu\nu}\Psi_{b\bt}
\right)~, \label{defA} \\
&& {\cal D}_{\bt}^j = {\partial \over \partial \theta^\bt_j} -{1\over 2} 
\;\theta^{j\al}\partial_{\al\bt}~,\qquad 
\{{\cal D}^i_\al, {\cal D}^k_\bt \} = 
\epsilon^{ik}\partial_{\al\bt}~. \label{defD}
\eea    
It is easy to find the full nonlinear algebra of the covariant derivatives 
$\nabla^i_\alpha, \nabla_{\rho\beta}$. We explicitly give the 
anticommutator of spinor derivatives
\be
\{\nabla^i_\alpha, \nabla^k_\beta \} \equiv -\nabla^{ik}_{\alpha\beta} 
= 
\left[{1\over 2}
(\delta^\omega_\alpha\delta^\sigma_\beta 
- \delta^\sigma_\alpha\delta^\omega_\beta)\epsilon^{ki} + 
\epsilon^{\omega\sigma\gamma\tau}(\nabla^i_\alpha\Psi^d_\gamma)
(\nabla^k_\beta\Psi_{d\tau}) \right]\nabla_{\omega\sigma}~. \label{nablalg}
\ee

We observe that, besides expressing Goldstone superfields through the only 
basic one $q^{ia}$, eq. \p{basconstr} imposes the nonlinear constraint 
\p{basconstr2} on this superfield. We recognize it as a nonlinear 
generalization of the well-known hypermultiplet constraint
\cite{fs} 
\be 
{\cal D}^{(i}_\bt\; q^{k)a} = 0~. \label{constrfree} 
\ee
The latter reduces the field content of $q^{ia}(x, \theta)$ to 
four bosonic and eight fermionic components      
\be
q^{ia}(x, \theta) \;\;\Rightarrow \;\; \phi^{ia}(x) + \theta^{\alpha i} 
\psi_\alpha^a (x) + x\mbox{-derivatives}~, \label{reduct}
\ee
and simultaneously puts these fields on shell
\be
\Box \phi^{ia}(x) = 0~, \quad \partial^{\alpha\beta}\psi^a_\bt = 0 \quad 
\left(\Box \equiv \partial^{\alpha\beta}\partial_{\alpha\beta} = {1\over 2} 
\;\epsilon^{\al\bt\mu\nu}\partial_{\alpha\beta}\partial_{\mu\nu}\right)~.
\ee
Eq. \p{basconstr2} is expected to yield a nonlinear generalization of
the $d=6$ hypermultiplet irreducibility conditions and equations of 
motion. It follows from \p{nablalg} that all superfields 
obtained by the successive action of $\nabla^i_\alpha$ on $\Psi_{a\beta}$ 
are reduced to ordinary $x$-derivatives of $q^{ia}$ and $\Psi_{a\beta}$, 
i.e. these two superfield projections indeed exhaust the irreducible 
fields content of $q^{ia}(x,\theta)$.  
 
Inspecting how the spontaneously broken nonlinear (super)symmetries  
\p{susy2} - \p{Ztr} are realized on the components of $q^{ia}$ 
(at the linearized 
level), we find that $\phi^{ia}(x)$ and $\psi_\alpha^a(x)$ are just 
Goldstone fields associated with the broken $Z$-translations and $S$- 
supertranslations, while the Goldstone fields accompanying the spontaneous
breakdown of the $SO(1,9)/ SO(1,5)\times SO(4)$ transformations,
$\partial_{\alpha\beta}\phi^{ia}(x)$, are recognized as the coefficients 
of the second-order $\theta$ monomials in the $\theta$-expansion 
of $q^{ia}(x,\theta)$.   

Thus the only essential Goldstone superfield supporting
the partial spontaneous breaking of $N=1,\; D=10$ SUSY down to 
$N=(1,0), \; d=6$ within the nonlinear realization scheme is 
the hypermultiplet 
superfield $q^{ia}(x, \theta)$. It is subjected to the nonlinear 
dynamical constraint \p{basconstr2} and accommodates all the Goldstone 
fields associated with the spontaneously broken symmetry generators 
including those of the $D=10$ Lorentz coset $SO(1,9)/ SO(1,5)\times SO(4)$. 

Note that the Lorentz Goldstone superfield 
$\Lambda^{ia}_{\al\bt}$  algebraically enters also 
into the Cartan form $\Omega_S \equiv \Omega_{S\;\al b}S^{\al b}$, 
$\Omega_{S\;\al b}= d\Psi_{\al b} - 2 \Lambda_{ib\gamma\al}
d\theta^{i\gamma} + \ldots $. This could mean that 
there exists an alternative way to eliminate $\Lambda^{ia}_{\al\bt}$, 
that time in terms of spinor derivative of $\Psi_{\al b}$ by equating 
to zero the appropriate part of the covariant $d\theta$-projection of  
$\Omega_S$. However, a careful analysis making use of the 
Maurer-Cartan equations shows that this part identically vanishes 
upon using the constraint \p{basconstr} (or eqs. \p{psieq}, 
\p{basconstr2}).          

It is worth mentioning that the kinematical and dynamical parts 
of eq. \p{basconstr} 
are separately covariant with respect to all hidden symmetries. 
In other words, eq. \p{basconstr2} is not implied by the formalism 
of nonlinear 
realizations, and should be regarded as a dynamical postulate. 
In the superembedding approach to superbranes 
\cite{{khark},{hs}} a similar postulate is known as 
``the geometro-dynamical principle'' or ``the basic constraint'' 
(see \cite{hs} and references therein). 
An interplay between 
the superembedding and PBGS approaches 
is discussed, e.g., in \cite{goteb}.  

To see which kind of dynamics 
is hidden in \p{basconstr2}, we considered it in the bosonic limit 
up to the first non-trivial 
order in fields, the third order. 
We found 
that it amounts to the following equation for $\phi^{ia}(x) \equiv 
q^{ia}(x,\theta)|_{\theta = 0}$
\be
\Box \phi^{ia} +{1\over 2}\;
\left(\partial_{\mu\nu}\phi \cdot 
\partial_{\rho\lambda}
\phi\right)
\partial^{\rho\lambda}\partial^{\mu\nu}\phi^{ia} = 0~, \label{NGeq}
\ee
where we omitted three-linear terms containing $\Box $ as they 
contribute to the next, 5th order, and used the notation 
$A\cdot B \equiv A^{ia}B_{ia} $. It is easy to see 
that eq. \p{NGeq} corresponds 
to the "static gauge" form of the bosonic $5$-brane Nambu-Goto 
(NG) action with the induced metric 
\be
g_{\rho\lambda\;\; \mu\nu} = 
{1\over 2}\;
\left( \epsilon_{\rho\lambda\mu\nu} 
- \partial_{\rho\lambda}\phi \cdot\partial_{\mu\nu}\phi \right) 
\equiv 
{1\over 2}\;
\left( \epsilon_{\rho\lambda\mu\nu} 
- d_{\rho\lambda\;\mu\nu}\right)~, 
\label{metrind} \ee  
that is  
\bea 
S_{NG} &=& \mbox{const}\;\int d^6x \;\left(\sqrt{-\mbox{det}\;g} 
-1\right) \nn 
&\sim & \int d^6 x \left\{ \mbox{Tr}\;d
-{1\over 8}\left(\mbox{Tr}\;d \right)^2 + {1\over 4}\;\mbox{Tr}\;d^{\;2} + 
O(d^{\;3}) \right\}. \label{NGact}
\eea 
Though it remains to prove that the higher-order 
corrections are combined into this nice geometric form, 
the above consideration suggests that this is very likely 
(in sect. 6 we show this on the simplified $d=1$ example). 
Then eq. \p{basconstr2} can be viewed as a 
manifestly $N=(1,0),\; d=6$ worldvolume superymmetric PBGS form of 
the equations of the scalar super 5-brane in $D=10$ \cite{aetw}(we use 
the nomenclature of ref. \cite{sezg}). So the nonlinear realization 
description of the partial breaking $N=1,\; D=10 \,\Rightarrow \, N=(1,0),\; d=6$ 
admits the natural brane interpretation, much in line of the previous 
studies \cite{hp} - \cite{gpr}. 
     
\vspace{0.3cm}
\noindent{\bf 5. Brane extension of harmonic Grassmann analyticity?}
For further discussion it will be convenient to project all the involved 
quantities on the $SU(2)$ harmonics $u^{\pm i}, u^{+i}u^-_i =1$ \cite{gikos} 
\be
\theta^\al_i \;\Rightarrow \; \theta^{\pm \al} = 
\theta^{\al i}u^\pm_i~, \;\;
\nabla^i_\al \;\Rightarrow \;\nabla^\pm_\al = \nabla^i_\al u^\pm_i~, 
\;\;q^{ia} \; \Rightarrow \; q^{\pm a} = q^{ia}u^\pm_i~. \label{proj}
\ee
Then the basic eq. \p{basconstr2} can be rewritten as  
\be
\nabla^+_\al q^{+a} = 0~.  \label{basconstr3} 
\ee
In the standard hypermultiplet case an analogous condition means that 
$q^{+a}$ lives on an analytic subspace of the full harmonic superspace 
$(x, \theta, u)$, and  this was the starting point 
of construction of off-shell actions for the hypermultiplet in \cite{gikos}.  

A difficulty with a similar treatment of \p{basconstr3} stems from the fact 
that the anticommutator of two $\nabla^+_\alpha $ is not vanishing, in 
contrast to its flat prototype 
\be
\{\nabla^+_\al, \nabla^+_\bt\} = 
-
\epsilon^{\rho\lambda\gamma\tau}(\nabla^+_\al\Psi^d_\gamma)\;
(\nabla^+_\bt\Psi_{d\;\tau}) \nabla_{\rho\lambda} \equiv - 
F^{++\rho\lambda}_{\;\;\;\al\bt}\;\partial_{\rho\lambda}~. \label{++} \\
\ee
As a result one has an extra integrability condition 
\be
F^{++\;\rho\lambda}_{\;\;\;\al\bt}\partial_{\rho\lambda} q^+_a = 0~, 
\label{intcond} 
\ee
which could be too strong (e.g., implying $q^{ia}$ to be a constant). 
We have checked that, up to the seventh order in $q^{ia}$, 
this condition is satisfied {\it identically} as a consequence 
of the structure of $F^{++\;\rho\lambda}_{\;\;\;\mu\nu}$. 
It is plausible that this holds to any order and in what follows 
we can take for granted that \p{intcond} produces no new 
restrictions on $q^{+a}$. 

Then eq. \p{basconstr3} implies, as usual, the existence 
of an analytic basis in the harmonic superspace where $\nabla^+_\al$ is 
reduced to the partial derivative with respect to $\theta^{-\nu}$ (when 
applied to $q^{+a}$) and where $q^{+a}$ lives as an unconstrained 
analytic superfield. The coordinate transformation to this basis
should be highly nonlinear in the involved fields. 

Instead of trying to find such a change of coordinates, it is  
easier to seek for a brane generalization of the standard off-shell 
$q^+$ action, i.e. for the action yielding in the bosonic sector the 
whole NG action \p{NGact}. The possibility that such an action exists
for the considered case was noticed in \cite{hs}. 
It is curious that 
there indeed exists a quartic extension of the standard free $q^+$ 
action which correctly reproduces the first terms in \p{NGact}. 
It reads    
\be
\tilde{S}_q \; \sim \; \int d\zeta^{(-4)} q^{+}_a {\cal D}^{++}q^{+a} + 
\alpha  \int d Z  
\;(q^{+}_a {\cal D}^{--}q^{+a})^2~.
\label{corr}
\ee
Here $d Z[du]$ and $d\zeta^{(-4)}$ are the appropriate 
integration measures over $d=6$ harmonic superspace and its analytic 
subspace, 
\be
dZ = d\zeta^{(-4)}({\cal D}^+)^4\,, \; d\zeta^{(-4)} = d^6 x [du] ({\cal D}^-)^4\,,\;\; 
({\cal D}^{\pm})^4 = {1\over 4!}\epsilon^{\alpha\beta\gamma\lambda}{\cal D}^{\pm}_\alpha
{\cal D}^{\pm}_\beta{\cal D}^{\pm}_\gamma{\cal D}^{\pm}_\lambda \;,  \label{defmeas}
\ee
${\cal D}^{\pm\pm} =  \partial^{\pm\pm} - 1/2\,
\theta^{\pm \al}\theta^{\pm \bt}\partial_{\al\bt} + 
\theta^{\pm \al}\partial/\partial \theta^{\mp \al}$ 
are harmonic derivatives, $\alpha $ is a 
dimensionless parameter (we use the same notation for the central-basis 
and analytic $q^{+a}$, hoping that this will not lead to confusion).
The first term in \p{corr} is the standard free $q^+$ action. 
We have found that after eliminating auxiliary fields 
(beyond expectation, they do not propagate) and making appropriate 
nonlinear redefinition of the physical bosonic field $\varphi^{ia}(x)$
($q^{+a}| = \varphi^{ia}u^+_i + \ldots $), 
$$
\varphi^{ia} = \phi^{ia} -{\al \over  24}\left\{[(\Box \phi \cdot \phi) 
- (\partial\phi\cdot \phi)]\,\phi^{ia} - 
{1\over 4}\,(\phi)^2\Box\phi^{ia} + {1\over 2}(\phi\cdot  
\partial_{\al\bt}\phi)\,\partial^{\al\bt}\phi^{ia} \right\} + 
O(\phi^5)~,
$$
the bosonic part of the component action in \p{corr} in fourth order 
in fields coincides with \p{NGact} under the choice $\alpha = -2/3$. 

This observation suggests the existence of the $q^+$ action with 
the whole static-gauge NG action in the bosonic sector. 
Clearly, the superfield equations of motion following from it, together with 
the analyticity condition, should amount to the basic nonlinear 
constraint \p{basconstr2} (or \p{basconstr3}). This action should be 
$N=(1,0),\;d=6$ ($N=2,\; d=4$) counterpart of the Goldstone chiral 
superfield action of ref. \cite{{bg1},{bg3}}. Possible existence of 
such a brane analog of the free off-shell $q^+$ action 
raises the question what could be brane analogs of 
$q^+$ actions with interaction. The latter yield 
hyper-K\"ahler sigma models in their bosonic sector. 
Presumably, their brane extensions could correspond to 
super 5-branes on non-trivial curved backgrounds.  

All such actions, being generalizations of off-shell 
$q^+$ actions, should necessarily involve infinite sets of auxiliary 
fields. They could provide an interesting alternative to the standard 
Green-Schwarz-type 
lagrangian description of superbranes \cite{hlp,aetw,gs}. It would be important 
to find the symmetry principles behind their structure. In the next section 
we present further evidence in favour of the existence of such actions.   

\vspace{0.3cm}
\noindent{\bf 6. $N=2\,,\;D=5$ superparticle.}
All the relations presented so far admit simple dimensional 
reduction to the $d=5$ and further $d=4~,...1$ worldvolumes 
by neglecting dependence on the corresponding worldvolume 
coordinates (in the Green-Schwarz approach this amounts to 
the ``double dimensional reduction''). 
One gets in this way manifestly worldvolume 
supersymmetric superfield equations of super 4-brane 
in $D=9$, super 3-brane in $D=8$, supermembrane in $D=7$ and so on, 
up to $N=2$ superparticle in $D=5$. They all have $8$ manifest and $8$ nonlinearly 
realized supersymmetries. Here we illustrate our consideration 
on the example of $N=2\,, \;D=5$ superparticle. 

In this case the basic anticommutation relations \p{susy} become
\be
\left\{ Q_{\al}^i,Q_{\bt}^j\right\}=\ep^{ij}\Omega_{\al\bt}P\; ,\quad
\left\{ Q_{\al}^i,S^{a\bt}\right\} = \de_{\al}^{\bt}Z^{ia}\; , \quad
\left\{ S^{a\al},S^{b\bt}\right\} = -\ep^{ab}\Omega^{\al\bt}P \;. 
\label{susy1d}
\ee   
The full automorphism group of 
\p{susy1d} is the product $Spin(1,4)\times Spin(5)$; the first 
factor is the target 
$D=5$ Lorentz group which acts on the indices $i, a$
\be 
SO(1,4) \sim Spin(1,4)\qquad \propto \quad \left\{T^{ij}, T^{ab}, 
K^{ia} \right\}\;, 
\ee
and $Spin(5)$ acts on the spinor indices.
In \p{susy1d}, $\Omega_{\al\bt} = - \Omega_{\bt\al}$
is the 
invariant $Spin(5)$ symplectic metric allowing to 
raise and lower the spinor indices 
($ \Omega^{\al\bt} = 
-{1\over 2}\epsilon^{\al\bt\rho\gamma}\Omega_{\rho\gamma}, \; 
\Omega_{\al\bt}\Omega^{\bt\gamma} = \delta^\gamma_\al$), $P$ 
is the worldline translations operator. 

Basically, the reduction to 
the case at hand is accomplished via the substitution $\partial_{\al\bt} = 
\Omega_{\al\bt}\partial_t$, where $t = \Omega_{\al\bt}x^{\al\bt} = 
- \Omega^{\al\bt}x_{\al\bt}$ is the worldline coordinate. 
The relations \p{psieq}, \p{basconstr2} 
(in the notation using $SU(2)$ harmonics) preserve their form, 
\be
\nabla^+_\al q^{+a} = 0~, \qquad \Psi^a_\bt =  \nabla^+_\bt q^{-a} = 
- \nabla^-_\bt  q^{+a}~, \qquad \tilde{\Lambda}^{\pm a} = E^{-1}
\partial_t q^{\pm a}~, \label{eqs1}
\ee 
with 
\bea
\nabla^{\pm}_\alpha &=& {\cal D}^{\pm}_\alpha - 
{1\over 2}\Psi^{b\bt}{\cal D}^{\pm}_\alpha \Psi_{b\bt}
E^{-1}\partial_t~, \quad
E = 1 - {1\over 2}\Psi^{a\al}\partial_t\Psi_{a\al}~, \nn
\{\nabla^+_\alpha, \nabla^-_\beta \} &=& \nabla^{+-}_{\al\bt} 
= F_{\al\bt}\partial_t~, 
\quad  F_{\al\bt} = -E^{-1}\left[ \Omega_{\al\bt} + \nabla^+_{\al}
\Psi^{b\rho}\nabla^-_{\bt}\Psi_{b\rho} \right]~. 
\eea 
Acting on $\Psi^a_\bt$ in \p{eqs1} by covariant derivatives, one finds 
\be
\nabla^+_\al \Psi^a_\bt = -F_{\al\bt}\partial_t q^{+a}~, 
\qquad
\nabla^-_\al \Psi^a_\bt = F_{\bt\al}\partial_t q^{-a}~, \label{eqs12}
\ee
whence it follows, in particular, that 
\be
\{ \nabla^+_\al, \nabla^+_\bt \} = 0~. \label{anal1}
\ee 

Looking at the matrix $F_{\al\bt}$, one observes that 
eqs. \p{eqs12} are the system of nonlinear equations for the 
unknowns  $\nabla^{\pm}_\al \Psi^a_\bt$. In the given simplified 
case it can be explicitly solved, and, further, the explicit 
expression for $\Psi^a_\bt$ in terms of $q^{\pm b}$ 
can be found. For our purposes it is enough to give the solution 
in the bosonic limit, with all fermions discarded
\be
\nabla^{\pm}_\al \Psi^a_\bt\; | = {\cal D}^{\pm}_\al \Psi^a_\bt   
= 2 \Omega_{\al\bt}\;\frac{1}{1 + \sqrt{1  - v^2}}\;\partial_t 
q^{\pm a}~, \quad (v^{ia} \equiv  \sqrt{2}\;\partial_t q^{ia})~.
\ee

The constraint in \p{eqs1} implies the following equation 
(once again, with all fermions omitted)
\be
\left[ \left(\nabla^{+-}_{\bt\nu}\nabla^{+-}_{\rho\gamma} + 
\nabla^{+-}_{\nu\gamma}
\nabla^{+-}_{\rho\beta} - \nabla^{+-}_{\bt\gamma}
\nabla^{+-}_{\rho\nu}\right) - 
\{\nabla^+_\rho, [\nabla^-_\gamma,  \nabla^{+-}_{\bt\nu}]\}\right]
q^{+}_a  
+ \{\nabla^+_\rho, [\nabla^+_\bt,  \nabla^{+-}_{\nu\gamma}]\}q^{-}_a = 0~.
\label{bos1}
\ee
A straightforward calculation shows that the terms 
with spinor derivatives in this relation identically  
vanish, while the term within the parenthesis 
yields, modulo an overall scalar factor, 
the dynamical equation for 
$q^{ia}(t)$ (we write it in terms $v^{ia} = \sqrt{2}\,\partial_tq^{ia}$) 
\be
\partial_t v^{ia}E_{ia}^{kb} \equiv 
\partial_t v^{ia}\left(I_{ia}^{kb} + 
\frac{\partial_t  v_{ia}\partial_t v^{kb}}{1 -v^2 +
\sqrt{1 - v^2}}\right) = 0~. \label{boseq1}
\ee
After multiplying from the 
right by the matrix $E$ and using 
\be
(E^2)^{ia}_{kb} = I^{ia}_{kb} +  
\frac{\partial_t  v^{ia}\partial_t v_{kb}}{
1 - v^2}~,
\ee
one rewrites \p{boseq1}, up to a scalar factor, in the form 
\be
\frac{d}{d t}\left(  \frac{v^{ia}}{\sqrt{1 - v^2}} \right) = 0~, 
\ee
that is recognized as the equation of motion corresponding to the 
static-gauge form of the NG action for the massive particle in $M^{1,4}$ 
\be
S \sim \int dt \sqrt{1 - v^2}~. \label{NG1d}
\ee

Although, due to the specificity of the $d=1$ case, the above bosonic 
equation actually amounts to the free one 
$\partial_t v^{ik} = 0$, \footnote{Nonetheless, 
the action corresponding to this equation should be  
just \p{NG1d}, because it is the unique bosonic action that respects
the nonlinearly realized $SO(1,4)/SO(4)$ hidden 
symmetry of the constraint for $q^{ia}$ in \p{eqs1}.} 
we expect that in the non-trivial $d>1$ cases the constraint \p{eqs1} 
yields the equation of motion for $q^{ia}$ in the form similar to 
\p{boseq1}, and it takes the standard NG form only after 
rotating the free target space index by an appropriate field-dependent 
non-degenerate matrix. Actually, when we performed the lowest-order 
computation outlined in sect. 4, we met just this peculiarity.

Finally, we address the issue of existence of the off-shell 
harmonic analytic action for this simplest system. Since 
in the present case the integrability condition \p{anal1} is valid generically  
(not only when applied on $q^{+a}$), the analytic basis 
definitely exists. Like in the $d=6$ case, we shall try to construct the action 
directly in the analytic harmonic $d=1$ superspace $\zeta = (t, \theta^{+\alpha}, 
u^{\pm i})$. We start from the $d=1$ reduction of the action \p{corr} (with 
$d^6 x \rightarrow dt$ in eq. \p{defmeas}) 
\bea 
S_q &\sim & S_{0} + S_{1} = \int d\zeta^{(-4)} q^{+}_a{\cal D}^{++}q^{+a} + 
\alpha \int d Z  \;A^2  \label{d14} \\
A &=& q^{+}_a{\cal D}^{--}q^{+a} \equiv q^{+} \cdot {\cal D}^{--}q^{+}\;. \label{defA1} 
\eea
We vary this action with respect to the lowest-order part of the broken 
SUSY transformation
\be
\delta_{(0)} q^+_a = c^{+}_a = \eta^\alpha_a\theta^+_\alpha \;.\label{zeroth}
\ee
The free part in \p{d14} is obviously invariant while 
the quartic part is not  
\be 
\delta_{(0)} S_1 = \alpha \int dZ  \left\{ L^{-3}\cdot{\cal D}^{++}q^+  + 
2A\; (q^+\cdot c^-)\right\}\;. \label{delta01}
\ee 
Here 
\be
L^{-3 a} = c^{-a}\; q^+\cdot({\cal D}^{--})^2q^+ +
 q^{+a}\; c^-\cdot ({\cal D}^{--})^2q^+ + 2 {\cal D}^{--}q^{+a}\; c^-\cdot {\cal D}^{--}q^+\,, 
\; c^-_a = \eta^\alpha_a \theta^-_\alpha\;. \label{defi}
\ee
The second term in \p{delta01} vanishes as a consequence of analyticity of $q^+$, while 
the first term can be cancelled by the appropriate analyticity-preserving 
variation of $q^+$ in the free part of the action  
\be 
\delta_{(1)} q^+_a = -\alpha ({\cal D}^+)^4 L^{-3}_a\;. \label{first}
\ee
Thus the first nonlinear term in the variation of $q^+$ under the hidden SUSY 
is also uniquely defined. Already at this step we observe an important 
phenomenon. Commuting $\delta_{(1)}$ with $\delta_{(0)}$, one immediately finds 
that, to the first order in $q^+$, the correct closure $\sim \partial_t q^+_a$ for the broken 
SUSY is achieved only modulo equations of 
motion. In other words, it is {\it impossible} to keep 
off shell both hidden and manifest SUSY's, the best we can gain is the off-shell world-line $N=8$ SUSY \footnote{This situation is quite similar 
to the formulation of $N=4\,, \;D=4$ super Yang-Mills theory via unconstrained 
harmonic $N=2\,, \; D=4$ superfields \cite{gios}: only the manifest 
$N=2$ SUSY is off-shell in such a formulation.}. 

The next steps in our recursion procedure are to 
compute the variation $\delta_{(1)} S_1$ and to look for the sixth-order 
correction $S_{2}$, such that $\delta_{(0)} S_2$ cancels 
$\delta_{(1)} S_1$. We proceed from the most general sixth-order Lagrangian 
density of the dimension $-4$, local in harmonics and having zero 
harmonic $U(1)$ charge. A part of its variation has the form $\sim {\cal D}^{++}q^+_a$ 
and hence can be cancelled by the appropriate shift of $q^+$ in the free action 
(it is of the fourth order in $q^+$). The remaining part is required to 
cancel $\delta_{(1)} S_1$. This requirement, together with that of on-shell closure of 
the hidden SUSY to the third order in $q^+$, uniquely (up to a freedom in the choice 
of independent structures in the Lagrangian) fix $S_2$ to be 
\bea
S_2 &=& {2\alpha^2 \over 5}\int dZ \left\{ 2\, A B^{++} ({\cal D}^{--}q^+\cdot 
{\cal D}^{--}\partial_t\, q^+) - {3\over 2}\, A({\cal D}^{--}B^{++})^2 
 \right. \nn 
&& \left.-\; 7\,{\cal D}^{--}A\, 
{\cal D}^{--}B^{++}\, B^{++} - {7\over 2}\, A\,\partial_t\,A\, {\cal D}^{--}B^{++} \right\}\;, 
\label{corr6}
\eea
where 
\be 
B^{++} = q^+\cdot \partial_t\,q^+\;.
\ee
To simplify this expression, let us treat $B^{++}$ as the
analytic potential of some composite $N=8\,,\;d=1$ (dimensionally 
reduced $N=(1,0)\,, \;d=6$) vector multiplet \cite{gikos} 
and introduce a non-analytic potential $B^{--}$ by the 
standard relation \cite{bz}
\be
{\cal D}^{--}B^{++} - {\cal D}^{++}B^{--} = 0\;.
\ee
We substitute it into \p{corr6}, make use of the identity 
$$
2B^{++} = {\cal D}^0 B^{++} = [{\cal D}^{++}, {\cal D}^{--}]B^{++} = 
({\cal D}^{++})^2 B^{--} - {\cal D}^{--}      
{\cal D}^{++} B^{++}\;, 
$$
and integrate by parts with respect to ${\cal D}^{++}$. In the course of this 
computation we omit all terms of the form $\sim ({\cal D}^{++}q^+\cdot F^{-3})$ 
as they can be absorbed into the redefinition of $q^+$ (the relevant shift 
is of the fifth order in $q^+$ and so 
does not affect $S_1$). The final answer for $S_2$ is as follows
\be
S_2 = 2\alpha^2 \int dZ\; A\, B^{--}\,B^{++}\;. \label{corr6fin}
\ee  

The existence of this sixth-order term is a non-trivial fact and can be 
regarded as a strong indication that the full harmonic action for this $D=5$ 
superparticle (and its higher-dimensional counterparts) exists. The form 
of $S_2$ \p{corr6fin} is rather suggestive: it looks like the 
harmonic superspace action of the composite $N=8\,, \;d=1$ vector multiplet $B^{++}$ 
in some background specified by the superfield $A$, both $B^{++}$ and $A$ 
being functions of the Goldstone hypermultiplet superfield $q^+_a$. 
This analogy could provide a hint of how to construct the full action. Also, it 
seems to imply a link with $N=8$ super D0-brane the worldline supermultiplet of which 
is just $N=8\,, \;d=1$ vector multiplet.                    
       
\vspace{0.3cm}
\noindent{\bf 7. Concluding remarks.} Besides the already mentioned 
problems for the future study, we list here a few other ones.

It is interesting to inquire whether some other $N=(1,0),\; d=6$ 
supermultiplets can be given the Goldstone interpretation and 
to which patterns of PBGS they could be relevant. The simplest one 
is the vector multiplet \cite{hst} comprising the fields $A_{[\mu\nu]}(x), 
\lambda^\mu_i(x), Y^{(ik)}(x)$. As the fermionic field $\lambda $ 
(a candidate for Goldstino) is of the same $d=6$ chirality as the 
Grassmann coordinate $\theta^{i\alpha}$, this multiplet 
can serve as the Goldstone one 
for the PBGS pattern $N=(2,0),\;d=6 \rightarrow N = (1,0),\; d=6$. By analogy 
with ref.\cite{bg2}, one can expect that the related theory is 
a manifestly $N=(1,0),\;d=6$ supersymmetric BI theory 
with the hidden nonlinearly 
realized rest of $N=(2,0),\;d=6$ SUSY (or $N=4$ BI theory with hidden 
extra $N=2$ SUSY in $D=4$). It is expected to yield 
a manifestly worldvolume supersymmetric 
PBGS description of super D5-brane in $D=6$ \footnote{These proposals were originally 
made in \cite{our}.}. 

The $N=(1,0)\,, \; d=6$ hypermultiplet parametrize transverse 
directions also in a special kind of super 5-brane in $D=10$, the heterotic 
5-brane obtained as a solitonic solution in the heterotic string theory \cite{str}. 
It was argued in \cite{witt2} that for quantum consistency of this solitonic 5-brane some extra worldsurface supermultiplets should be added, in particular, 
an $SU(2)$ gauge vector $N=(1,0)\,\;d=6$ multiplet. It would be interesting 
to understand the necessity of such additional $d=6$ multiplets within 
the PBGS approach. 
Note that the simple scalar super 5-brane to which our 
attention was limited here corresponds to another 
solution to the equations of the heterotic 
string theory, the ``neutral solution'' \cite{chs}.
        
It is intriguing to examine from the PBGS point of view $N=1,\;D=11$ 
(or the type IIA $N=2,\; D=10$) SUSY. The $N=(1,0),\; d=6$ superfield 
framework is suitable for studying the $1/4$ breaking of this SUSY. 
Let us see what happens in the linearized approximation. 

{}From the $d=6$ point of view, 
promoting $N=1,\;D=10$ SUSY to $D=11$ amounts to adding one more 
bosonic translation generator  $P_{11}$, two supertranslation 
generators of opposite chiralities $Q^{i\alpha}, S^a_\beta$ and two 
extra  Lorentz generators $U^{ia}$ and $W_{\al\bt}= - W_{\bt\al}$. 
The latter  
extend $SO(1,9)$ to $SO(1,10)$ and belong to the 
cosets $SO(5)/SO(4)$ and $SO(1,6)/SO(1,5)$. We still wish to have 
$N=(1,0),\; d=6$ SUSY as the only unbroken one, so we should add to 
the already incorporated Goldstone superfields several new ones associated 
with the extra generators 
\be
P_{11}\, \Rightarrow \,\Phi(x,\theta)\,, \, Q^{i\alpha} \,\Rightarrow 
\, \eta_{i\al}(x,\theta)\,, \, S^a_\al \,\Rightarrow \, 
\xi_a^\al (x,\theta)\,, \,
U^{ia} \,\Rightarrow \, u_{ia}(x,\theta)\,, \,
W_{\al\bt} \,\Rightarrow \,v^{\al\bt}(x,\theta)\,. \label{newG}
\ee
At the linearized level, the standard coset techniques yield the following 
expressions for the covariant $d\theta$-projections of the Cartan 1-forms 
related to the newly introduced (super)translations generators 
\be
P_{11}\,\Rightarrow \, {\cal D}^i_\al \Phi + \eta^i_{\al}~, \quad
Q^{i\al} \,\Rightarrow \,{\cal D}^i_\al \eta_{j\bt} + 
2\delta^i_j v_{\al\bt}~, \quad
S^a_\al \,\Rightarrow \, {\cal D}^i_\al \xi^\bt_a - \delta^\al_\bt u^i_a~.
\label{newC} 
\ee
One observes that 
the Goldstone superfields $\eta^i_a$, 
$v_{\al\bt}$, $u^{ia}$ (like $\Lambda^{ia}_{\al\bt}$ and $\Psi^a_\bt$) 
can be covariantly eliminated by equating to zero appropriate   
parts of the above projections of Cartan forms.      
On the other hand, the superfields $\Phi, \xi^\al_a$ can be shown to never 
appear linearly (without derivatives on them) in any Cartan form.   
So in the given case the set of unremovable Goldstone superfields enlarges 
to $\{ q^{ia}, \Phi, \xi^\al_a \}$. New superfields are reducible and 
we should impose on them proper constraints similar to the 
constraint  \p{basconstr2} for $q^{ia}$.  By analogy with the $D=10$ case we 
assume that the covariant elimination of the redundant Goldstone superfields 
and imposing constraints on the essential ones are simultaneously 
effected by  equating  to zero full $d\theta $-projections \p{newC} of 
the translation  and supertranslation Cartan forms (or 
the covariant nonlinear versions of \p{newC} in the full 
nonlinear case). As the result of such a procedure at the considered 
linearized level one gets the following  expressions for the new 
redundant Goldstone superfields
\be
v_{\al\bt} = -{1\over  4}\partial_{\al\bt}\Phi~, \qquad 
\eta^i_\al = -{\cal  D}^i_\al \Phi~, \qquad u^i_a = {1\over 4}
{\cal D}^i_\bt\xi^\bt_a~, \label{expr11}
\ee
and, simultaneously, the following constraints for the new unremovable ones
\be
\mbox{(a)} \quad  {\cal D}^{(i}_\bt{\cal D}^{k)}_\al \Phi = 0~; \qquad 
\mbox{(b)} \quad {\cal D}^{i}_\bt\xi^\al_a - {1\over 4}\,\delta^\al_\bt \, 
{\cal D}^{i}_\gamma\xi^\gamma_a = 0~. \label{constr11}
\ee     

The constraint (\ref{constr11}a) is immediately recognized 
as the one defining 
the self-dual tensor $N=(1,0),\; d=6$ supermultiplet in the field-strength 
formulation \cite{sok}. This constraint leaves 
the Goldstone  fermion 
$\eta^{i}_\al(x)$, a  scalar $\phi(x) = \Phi |$ (it parametrizes the 
broken eleventh direction) and a self-dual field strength 
$F_{(\al\bt)}(x)$ 
as the only irreducible fields in $\Phi(x,\theta)$ and puts all 
them on shell. The Goldstone superfields 
$q^{ia}$, $\Phi$ are naturally unified into a $N=(2,0)$ 
self-dual multiplet which is known to be 
the worldvolume multiplet 
of the M5-brane \cite{{hs},{gs},{m5}}. 
This nicely matches with the fact 
that these two $N=(1,0)$ multiplets 
realize the $1/2$ spontaneous breaking of $N=1,\;D=11$ SUSY 
down to $N=(2,0),\;d=6$ SUSY $\propto 
\{Q^i_\al, S^a_\bt, P_{\al\bt}, so(1,5)\oplus so(4)\}$. 

It is the remaining Goldstone superfield $\xi^\al_a$ which 
executes further breaking of this $N=(2,0),\;d=6$ SUSY down to 
$N=(1,0)$. Surprisingly, the constraint (\ref{constr11}b)
turns out to be too strong: it reduces $\xi^\al_b(x,\theta)$ 
to a few bosonic and fermionic constants
\bea
&& \mbox{eq.(47b)}  \quad \Rightarrow \quad 
\xi^\al_b (x, \theta) = \xi^\al_b(x) - \theta^{\al k}\;u_{kb} -{1\over 2} 
(\theta^{\al i}\theta^\bt_i)\;\phi_{\bt b}~, \label{solut} \\
&& \partial_{\al\bt}\,u_{bi} = 
\partial_{\al\bt}\,\phi_{b\gamma} = 0~, \quad 
\partial_{\gamma\bt}\,\xi^\al_b(x) + \delta^\al_\bt \phi_{\gamma b} - 
\delta^\al_\gamma \phi_{\bt b} = 0 \,\Rightarrow \,
\xi^\al_b (x) = \xi^\al_b + 2\,x^{\al\gamma}\phi_{\gamma b}~.\label{solut2}
\eea    
Nevertheless, this constraint is  
the only one which {\it (i)} is linear in ${\cal  D}^i_\al$ and {\it (ii)}
enjoys all linearly realized  symmetries.
We still do not know how to interpret this. 
Possible ways out are, e.g., to impose some alternative constraint 
of higher order in derivatives, or to retain the linearity in 
${\cal D}^i_\al$ but to allow an explicit breaking 
of the $D=11$ Lorentz symmetry and, simultaneously, 
of manifest $SO(4)$ symmetry, say, down to the diagonal 
$SU(2)$ subgroup. In this case there arises a possibility 
to impose on $\xi^\al_b$ 
the constraints identifying it with a superfield strength 
of $N=(1,0),\;d=6$ Maxwell multiplet \cite{hst} (they can be chosen 
on- or  off-shell). Of course, there remains a difficult problem 
of correct generalization to the full nonlinear case \cite{bg2}. 

Curiously, the constants in \p{solut}, \p{solut2} 
have true conformal dimensions and index structure 
for being parameters of some 
specific coset of superconformal extension of the $N=(2,0),\; d=6$ 
super Poincar\'e group, the supergroup $OSp(6,2|4)$ \cite{m5}.  
Indeed, $\xi^\al_b$ are going to be the parameters of the 
second Poincar\'e supertranslations, 
$u_{ia}$ the parameters of the coset $SO(5)/SO(4)$ and $\phi_{\gamma b}$ 
the parameters of one of two special supersymmetries.   

It is interesting to analyze from a similar standpoint also  
the type IIB $N=2, \; D=10$ SUSY. It can be argued that 
its $1/2$ breaking 
should be realized on the $N=(1,0)$ hypermultiplet  
and $N=(1,0)$ Maxwell field strength superfields 
as the Goldstone ones. Together they form an on-shell $N=(1,1)\,, \; d=6$ 
Maxwell-Goldstone multiplet. Further breaking to $N=(1,0),\;d=6$ SUSY 
in this case requires an extra essential fermionic Goldstone 
$N=(1,0)$ superfield $\nu^b_\al (x,\theta)$ constrained in an appropriate way. 
We failed to find a proper candidate for such superfield and 
constraints among the known $N=(1,0)\,, \;d=6$ multiplets.       

Finally, it is desirable to further clarify 
the relationships between the PBGS and superembedding approaches. 
They seem to be complementary to each other. The PBGS approach deals from 
the beginning with a minimal set of Goldstone superfields accommodating 
the physical brane degrees of freedom and it offers systematic 
techniques to deduce the transformation laws of these superfields 
under hidden nonlinear symmetries.  
On the other hand, superembedding approach allows one to classify 
physical worldvolume supermultiplets related to various superbranes 
and, under some assumptions (e.g., ``geometro-dynamical principle''), 
to learn whether these multiplets are on- or off-shell.  
In particular, the linearized 
analysis of the $N=1,\; D=10$ super 5-brane in ref. \cite{hs} 
(in the framework of conventional superspace) picks out just the $d=6$ 
hypermultiplet as a physical multiplet and predicts it to be on-shell.          
  
\vspace{0.4cm}
\noindent{\bf Acknowledgments.} We thank I. Bandos, E. Bergshoeff, 
F. Delduc, B. de Wit, R. Kallosh, S. Ketov, S. Kuzenko, O. Lechtenfeld, 
A. Pashnev, M. Tonin, A. van de Ven, M. Vasiliev and, especially, 
D. Sorokin for their interest in the work and 
useful comments. This research 
was supported in part by the Fondo Affari Internazionali Convenzione 
Particellare INFN-JINR. E.I. and S.K. acknowledge partial 
support from the grants RFBR 96-02-17634, RFBR-DFG 96-0200180, 
INTAS-93-127ext, INTAS-96-0538 and INTAS-96-0308.


\vspace{0.3cm}
\begin{thebibliography}{99}
\bibitem{1} S. Coleman, J. Wess, B. Zumino, Phys. Rev. {\bf 177} (1969) 
2239;\\
C. Callan, S. Coleman, J. Wess, B. Zumino, Phys. Rev. {\bf 177} (1969)
2247
\bibitem{2} D.V. Volkov, Sov. J. Part. Nucl. {\bf 4} (1973) 3
\bibitem{3} V.I. Ogievetsky, Proceedings of X-th Winter School of Theoretical 
Physics in Karpacz, Vol.1. p. 227 (Wroclaw, 1974)
\bibitem{bw} J. Bagger, J. Wess, Phys. Lett. {\bf B 138} (1984) 105
\bibitem{hp} J. Hughes, J. Polchinski, Nucl. Phys. {\bf B 278} (1986) 147
\bibitem{hlp} J. Hughes, J. Liu, J. Polchinski, Phys. Lett. {\bf B 180} 
(1986) 370                                                              
\bibitem{town} A. Achucarro, J. Gauntlett,  K. Itoh, P.K. Townsend, 
Nucl. Phys. {\bf B 314} (1989) 129
\bibitem{iak} E. Ivanov, A. Kapustnikov, Phys. Lett {\bf B 252} (1990) 439, 
{\bf B 267} (1991) 541E; {\it ibid} {\bf B 267} (1991) 175; 
Int. J. Mod. Phys. {\bf A 7} (1992) 
2153 
\bibitem{bg1} J. Bagger, A. Galperin, Phys. Lett. {\bf B 336} (1994) 25
\bibitem{bg2} J. Bagger, A. Galperin, Phys. Rev. {\bf D 55} (1997) 1091
\bibitem{bg3} J. Bagger, A. Galperin, Phys. Lett. {\bf B 412} (1997) 296
\bibitem{goteb} T. Adawi, M. Cederwall, U. Gran, M. Holm, B.E.W. Nilsson, 
Int. J. Mod. Phys. {\bf A 13} (1998) 4691
\bibitem{gpr} M. Ro\u{c}ek, A. Tseytlin, {\tt hep-th/9811232};\\
F. Gonzalez-Rey, I.Y. Park, M. Ro\u{c}ek, 
{\tt hep-th/9811130}
\bibitem{vak} D.V. Volkov, V.P. Akulov, Phys. Lett. 
{\bf B 46} (1973) 109
\bibitem{witten} E. Witten, Nucl. Phys. {\bf B 188} (1981) 513
\bibitem{gikos} A. Galperin, E. Ivanov, S. Kalitzin, V. Ogievetsky, 
E. Sokatchev, Class. Quant. Grav. {\bf 1} (1984) 469
\bibitem{fs} P. Fayet, Nucl. Phys. {\bf B 113} (1976) 135; \\
M.F. Sohnius, Nucl. Phys. {\bf B 138} (1978) 109
\bibitem{hst} P.S. Howe, G. Sierra, P.K. Townsend, Nucl. Phys. 
{\bf B 221} (1983) 331
\bibitem{hsw} P.S. Howe, K.S. Stelle, P.C. West, Class. Quant. Grav. {\bf 
2} (1985) 815 
\bibitem{bz} B.M. Zupnik, Sov. J. Nucl. Phys. {\bf 44} (1986) 512; 
Phys. Lett. {\bf B 183} (1987) 175
\bibitem{invh} E.A. Ivanov, V.I. Ogievetsky, Teor. Mat. Fiz. {\bf 25} 
(1975) 164  
\bibitem{khark} D. Sorokin, V. Tkach, D.V. Volkov, Mod. Phys. Lett. 
{\bf A 4} (1989) 901; \\
I. Bandos, P. Pasti, D. Sorokin, M. Tonin, D. Volkov,
Nucl. Phys. {\bf B 446} (1995) 79;\\
I.A. Bandos, D. Sorokin, D. Volkov, Phys. Lett. {\bf B 352} (1995) 269
\bibitem{hs} P.S. Howe, E. Sezgin, Phys. Lett. {\bf B 390} (1997) 133;\\
P.S. Howe, E. Sezgin, P.C. West, {\tt  hep-th/9705093}
\bibitem{aetw} A. Achucarro, J.M. Evans, P.K. Townsend, D.L. Wiltshire, 
Phys. Lett. {\bf B 198} (1987) 441
\bibitem{sezg} E. Sezgin, {\tt hep-th/9809204}
\bibitem{gs} I. Bandos, K. Lechner, A. Nurmagambetov, P. Pasti, D. Sorokin, 
M. Tonin, Phys. Rev. Lett. {\bf 78} (1997) 4332; \\
M. Aganagic, J. Park, C. Popescu, J.H. Schwarz, Nucl. Phys. {\bf B 496} 
(1997) 191; \\
P.S. Howe, O. Raetzel, E. Sezgin, J. High Energy Phys. 
{\bf 9808} (1998) 011;\\
P.S. Howe, O. Raetzel, I. Rudychev, E. Sezgin, {\tt hep-th/9810081}
\bibitem{gios} A. Galperin, E. Ivanov, V. Ogievetsky, 
E. Sokatchev, Class. Quant. Grav. {\bf 2} (1985) 617  
\bibitem{our} S. Bellucci, E. Ivanov, S. Krivonos, Presented at 
the XI Blokhintsev Conference (Dubna, July 13-17, 1998) and 
32nd Symposium Ahrenshoop (Buckow, September 1-5, 1998), to appear
in the Proceedings, {\tt hep-th/9809190} 
\bibitem{str} A. Strominger, Nucl. Phys. {\bf B 343} (1990) 167
\bibitem{witt2} E. Witten, Nucl. Phys. {\bf B 460} (1996) 541
\bibitem{chs} C.G. Callan, J.A. Harvey, A. Strominger, 
Nucl. Phys. {\bf B 359} (1991) 611; {\bf B 367} (1991) 60
\bibitem{sok} E. Sokatchev, Class. Quant. Grav. {\bf 5} (1988) 1459
\bibitem{m5} R. Kallosh, Phys. Rev. {\bf D 57} (1998) 3214; \\
P. Klaus, R. Kallosh, A. Van Proeyen, Nucl. Phys. {\bf  B 518} 
(1998) 117

\end{thebibliography}
\end{document}